\documentclass[aps,prd,reprint,groupedaddress,amsmath,amssymb,amsfonts,showpacs]{revtex4-1}

\usepackage{graphicx}
\usepackage{bm}
\usepackage{calc}
\usepackage{color}
\usepackage{wrapfig}
\usepackage{url}
\usepackage{paralist}
\usepackage{array}
\usepackage{tabularx}

\usepackage{txfonts}
\def\araa{ARA\&A}%
\def\apj{ApJ}%
\def\apjl{ApJ}%
\def\aap{A\&A}%
\def\jcap{J. Cosmology Astropart. Phys.}%
\def\mnras{MNRAS}%
\def\prd{Phys.~Rev.~D}%
\def\physrep{Phys.~Rep.}%
\newcommand{\sgn}{\operatorname{sgn}}

\begin{document}


\title{Revisiting metric perturbations in tensor-vector-scalar theory}


\author{Martin Feix}
\email[Electronic address: ]{feix@iap.fr}
\affiliation{CNRS, UMR 7095 \& UPMC, Institut d'Astrophysique de Paris, 98 bis Boulevard Arago, 75014, Paris, France}
\affiliation{Department of Physics, Technion - Israel Institute of Technology, Technion City, 32000 Haifa, Israel}

\date{\today}

\begin{abstract}
I revisit cosmological perturbations in Bekenstein's tensor-vector-scalar theory (TeVeS). Considering only scalar modes in the
conformal Newtonian gauge, the extra degrees of freedom are expressed in a way suitable for studying modifications at the level
of the metric potentials. Assuming a universe in the matter-dominated phase, I discuss the mechanism responsible for boosting
structure growth, and confirm the vector field as its key ingredient. Using a semi-analytic approach, I further characterize
the evolution of density perturbations and the potentials on sub- and superhorizon scales.
\end{abstract}


\maketitle


\section{Introduction}
\label{section1}
Despite the remarkable success of the standard cosmological model \cite{largescale, *Benjamin2007, *Guy2010, *Planck2015}, open
questions related to the nature of cosmic acceleration and dark matter \cite{Bertone2005, *Frieman2008} still leave room to ponder
on alternative possibilities. Among the various proposals, it has been suggested that the gravitational dynamics may be different
from the predictions of general relativity (GR). Additionally motivated by considerations beyond the field of cosmology, this has
triggered theoretical developments aimed at finding viable models, and there exists now a large number of modified theories which
typically introduce extra degrees of freedom in the gravitational sector \cite{Clifton2012}.

The phenomenology of such modified gravity theories is quite rich and can involve complex and non-intuitive results which are hard to
understand from the equations of motion alone. A way forward is to use appropriate parameterizations which capture the new fields
and their dynamics, but allow one to view their effects from a different angle and to explore how familiar quantities and observables
are affected. Here I want to follow this approach using the example of Bekenstein's tensor-vector-scalar theory (TeVeS) \cite{teves}
which was originally constructed as a relativistic extension for the modified Newtonian dynamics (MOND) paradigm \cite{Mond1, *Mond3,
*mondnew} and has been subject to numerous studies in the literature \cite{Famaey2012}. Considering perturbations around a spatially
flat Friedmann-Robertson-Walker (FRW) background, I will express the additional degrees of freedom in a way suitable for studying
modifications at the level of the metric potentials in the conformal Newtonian gauge. This particular ansatz will then be used to
study several aspects of the cosmological evolution such as the modified growth of density perturbations and the gravitational slip
which has been identified as a generic feature of modified gravity theories \cite{Bert2006, *Ferreira2010, *Bertschinger2011}. To
allow an analytic treatment of the cosmological background, I will adopt an Einstein-de Sitter (EdS) universe which only contains
pressureless matter and provides an excellent approximation to the matter era of a realistic universe. The approach taken in this work
differs in motivation from the recent attempts of parameterizing deviations from GR in a model-independent fashion \cite{Baker2011,
*Zuntz2011, *Baker2011b} which are interesting in their own right.

The paper is organized as follows: Starting with a brief review of TeVeS in Sec. \ref{section2}, its background cosmology and the
modified EdS model are presented in Sec. \ref{section3}. Considering perturbations in the conformal Newtonian gauge, I then introduce
a parameterization of the new variables and discuss the evolution of cosmological perturbations in Sec. \ref{section4}. Finally, I
conclude in Sec. \ref{section5}. For clarity, some of the material involving lengthy expressions is presented in an appendix. If not
stated otherwise, I will assume the notation of Ref. \cite{Skordis2006} throughout this work.

\section{Fundamentals of TeVeS}
\label{section2}
\subsection{Fields and action}
\label{section21}
In its original form, TeVeS \cite{teves,tevesreview} is a bimetric gravity theory which is based on three dynamical fields:
an Einstein metric $\tilde{g}_{\mu\nu}$, a time-like vector field $A_{\mu}$ such that
\begin{equation}
\tilde{g}^{\mu\nu}A_{\mu}A_{\nu} = -1,
\label{eq:3}
\end{equation}
and a scalar field $\phi$. The gravity-matter coupling involves a second metric $g_{\mu\nu}$ which is obtained from
\begin{equation}
g_{\mu\nu} = e^{-2\phi}\tilde{g}_{\mu\nu}-2A_{\mu}A_{\nu}\sinh(2\phi).
\label{eq:4}
\end{equation}
The frames delineated by the metric fields $\tilde{g}_{\mu\nu}$ and $g_{\mu\nu}$ are called {\it Einstein frame} and {\it matter frame},
respectively. The geometric part of the action is the same as in GR:
\begin{equation}
S_{g} = {\frac{1}{16\pi G}}\int \tilde{g}^{\mu\nu}\tilde{R}_{\mu\nu}\sqrt{-\tilde{g}}d^{4}x,
\label{eq:5}
\end{equation}
where $\tilde{R}_{\mu\nu}$ is the Ricci tensor of $\tilde{g}_{\mu\nu}$ and $\tilde{g}$ the determinant of $\tilde{g}_{\mu\nu}$. The
vector field's action $S_{v}$ reads as follows:
\begin{equation}
S_{v} = -\frac{1}{32\pi G}\int\left\lbrack K_{B}F^{\mu\nu}F_{\mu\nu}-\lambda(A_{\mu}A^{\mu}+1)\right\rbrack\sqrt{-\tilde{g}}d^{4}x,
\label{eq:6}
\end{equation}
where $F_{\mu\nu} = \tilde{\nabla}_{\mu}A_{\nu}-\tilde{\nabla}_{\nu}A_{\mu}$ and indices are raised and lowered with respect to
$\tilde{g}_{\mu\nu}$, i.e. $A^{\mu}=\tilde{g}^{\mu\nu}A_{\nu}$. Here the constant $K_{B}$ describes the coupling of the vector field to
gravity and $\lambda$ is a Lagrangian multiplier enforcing the normalization condition of $A_{\mu}$. The action $S_{s}$ of the scalar
field $\phi$ involves an additional non-dynamical scalar field $\mu$, and takes the form
\begin{equation}
S_{s} = -\frac{1}{16\pi G}\int\left\lbrack\mu h^{\mu\nu}\tilde{\nabla}_{\mu}\phi\tilde{\nabla}_{\nu}\phi+V(\mu)\right\rbrack\sqrt{-\tilde{g}}d^{4}x,
\label{eq:7}
\end{equation}
where $h^{\mu\nu} = \tilde{g}^{\mu\nu}-A^{\mu}A^{\nu}$ and $V(\mu)$ is an initially arbitrary (potential) function. As the field $\mu$
is related to the invariant $h^{\mu\nu}\tilde{\nabla}_{\mu}\phi\tilde{\nabla}_{\nu}\phi$, however, it could in principle be eliminated
from the action. Finally, the matter action is given by
\begin{equation}
S_{m} = \int\mathcal{L}_{m}\left\lbrack g,\Upsilon^{B},\nabla\Upsilon^{B}\right\rbrack\sqrt{-g}d^{4}x,
\label{eq:8}
\end{equation}
where $\Upsilon^{B}$ is a generic collection of matter fields. By construction, world lines are geodesics of $g_{\mu\nu}$ rather than
$\tilde{g}_{\mu\nu}$. Despite its explicit bimetric construction, TeVeS may be written in pure tensor-vector form \cite{tv1} and
provides a particular example of general Aether-type theories \cite{tv2}.

\subsection{Choice of the potential, quasistatic systems\protect\\ and relation to MOND}
\label{section22}
The behavior of TeVeS in the nonrelativistic limit depends on the assumed potential $V$. Originally, Bekenstein made the choice
\begin{equation}
V(\mu) = \frac{3\mu_{0}^{2}}{128\pi l_{B}^{2}}\left\lbrack\hat{\mu}\left (4+2\hat{\mu}-4\hat{\mu}^{2}+\hat{\mu}^{3}\right )+2\log{(1-\hat{\mu})^{2}}\right\rbrack,
\label{eq:9}
\end{equation}
where the constant $l_{B}$ corresponds to a length scale, $\hat{\mu}=\mu/\mu_{0}$ and $\mu_{0}$ is a dimensionless constant \footnote{
Note the change of notation: Here $\mu$ is related to Bekenstein's auxiliary scalar field $\sigma_{B}$ through $\mu = 8\pi G\sigma_{B}
^{2}$ and the coupling constant of the scalar field, i.e. Bekenstein's $k_{B}$, is redefined in terms of $\mu_{0}$ using $\mu_{0} = 8
\pi/k_{B}$.}. GR is then recovered in the limit $K_{B}\rightarrow 0$ and $l_{B}\rightarrow\infty$. Applying the usual approximations
for weak fields and quasistatic systems, one finds $V^{\prime} \equiv dV/d\mu < 0$, and therefore $0< \mu < \mu_{0}$. Using that also
$V(\mu) < 0$ for the given range, the metric $g_{\mu\nu}$ turns out to be identical to the metric obtained in GR if the nonrelativistic
gravitational potential is replaced by
\begin{equation}
\begin{split}
W &= \Xi\Phi_{N}+\phi ,\\
\Xi &= e^{-2\phi_{C}}\left (1+K_{B}/2\right )^{-1},
\end{split}
\label{eq:10}
\end{equation}
where $\phi_{C}$ is the cosmological value of $\phi$ at the time the system in question breaks away from the cosmological expansion,
and $\Phi_{N}$ is the Newtonian potential generated by the matter density $\rho$ \footnote{For $\phi_{C}\neq 0$, the metric $g_{\mu
\nu}$ is not asymptotically Minkowskian. However, this is easily remedied by an appropriate rescaling of coordinates.}. In this
approximation, one further has
\begin{equation}
h^{\mu\nu}\tilde{\nabla}_{\mu}\phi\tilde{\nabla}_{\nu}\phi \rightarrow (\bm{\nabla}\phi)^{2} \equiv \lVert\bm{\nabla}\phi\rVert_{2}^{2},
\label{eq:11}
\end{equation}
and the equation of the scalar field reduces to
\begin{equation}
\bm{\nabla}\cdot\left (\mu\bm{\nabla}\phi\right ) = 8\pi G\rho.
\label{eq:12}
\end{equation}
Equations \eqref{eq:10} and \eqref{eq:12} correspond to the MOND paradigm \cite{teves}. If $\mu\rightarrow\mu_{0}$, the theory reaches
its Newtonian limit, and the measured gravitational constant $G_{N}$ is given by
\begin{equation}
G_{N} = \frac{\mu_{0} + 2 - K_{B}}{\mu_{0}(1-K_{B}/2)}G.
\label{eq:13}
\end{equation}
Similarly, the theory reaches its MONDian limit as $\mu\rightarrow 0$ and the acceleration constant $a_{0}$ can be expressed in terms
of the TeVeS and potential parameters,
\begin{equation}
a_{0} = \frac{\sqrt{6}}{2l_{B}}\frac{e^{\phi_{C}}}{\sqrt{\pi\mu_{0}}}\frac{G}{G_{N}}.
\label{eq:14}
\end{equation}
As can be seen from above, $a_{0}$ depends on $\phi_{C}$ and may therefore, in principal, change with time \cite{Bekenstein2008}.

\begin{figure}
\includegraphics[width=0.95\linewidth]{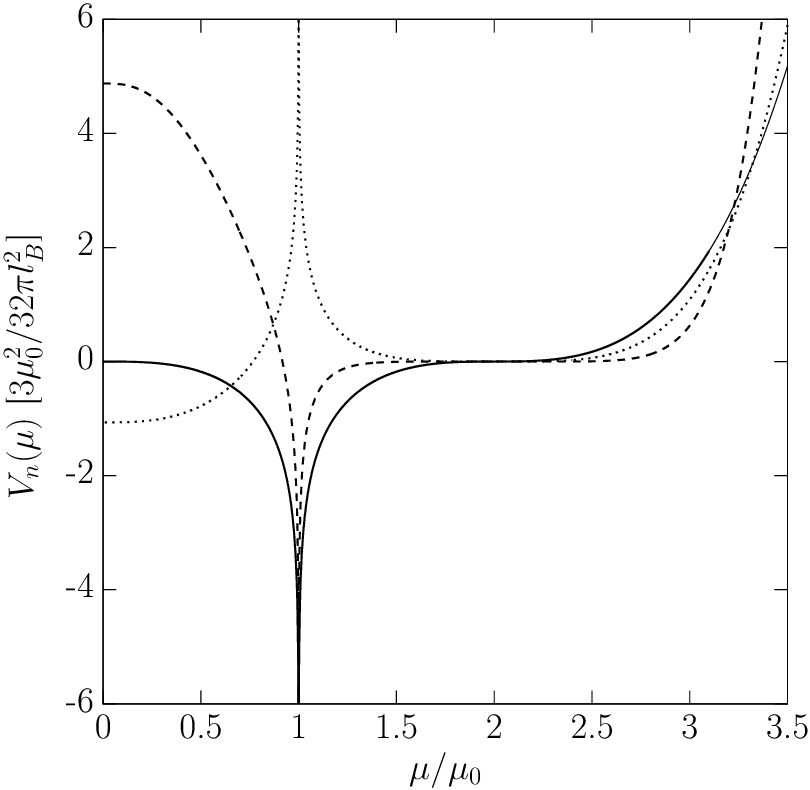}
\caption{Illustration of the generalized potential function $V_{n}(\mu)$ given by Eq. \eqref{eq:15} for $n=2$ (solid line), $3$
(dotted line), and $6$ (dashed line).}
\label{fig1}
\end{figure}

Different choices of $V$ and their implications for bound structures have been studied in Refs. \cite{freefunc, tevescosmo, Bourliot2007}.
Focusing on cosmological scales, I will assume the following class of potentials in this paper \cite{Bourliot2007}:
\begin{equation}
\begin{split}
V_{n}(\mu) &= \frac{3\mu_{0}^{2}}{32\pi l_{B}^{2}}\left\lbrack\frac{n+4+(n+1)\hat{\mu}}{(n+1)(n+2)}\left (\hat{\mu}-2\right )^{n+1}\right.\\
&+ \left.\frac{(-1)^{n}}{2}\log{(1-\hat{\mu})^{2}}+\sum\limits_{m=1}^{n}\frac{(-1)^{n-m}}{m}\left (\hat{\mu}-2\right )^{m}\right\rbrack ,
\end{split}
\label{eq:15}
\end{equation}
where $n\geq 2$ \footnote{This class of potentials will modify the dynamics of quasistatic systems if $n\neq 2$, but this can be
avoided by resorting to a piecewise definition of $V(\mu)$.}. Adopting different values of $n$, Fig. \ref{fig1} illustrates the
resulting potential shape as a function of $\hat{\mu}$. The such generalized potential reduces to Bekenstein's toy model if $n=2$.
The derivative of $V_{n}(\mu)$ takes a simpler form and can be expressed as
\begin{equation}
V_{n}^{\prime}(\mu) = \frac{3\mu_{0}}{32\pi l_{B}^{2}}\hat{\mu}^{2}\frac{\left (\hat{\mu}-2\right )^{n}}{\hat{\mu}-1}.
\label{eq:16}
\end{equation}
Requiring that $V^{\prime}$ is single-valued and $V^{\prime}\ge 0$ \cite{teves}, one is always free to choose between two possible
potential branches. Here I shall use the branch ranging from the extremum at $\mu =2\mu_{0}$ to infinity. Under these preliminaries,
the potential in Eq. \eqref{eq:15} gives rise to tracker solutions of the scalar field \cite{Bourliot2007}, with a background evolution
similar to other theories involving tracker fields \cite{Ratra1988, *Ferreira1997, *Skordis2002}. I will further elaborate on this
behavior in Sec. \ref{section32}.

\section{Cosmological background}
\label{section3}
\subsection{Evolution equations}
\label{section31}
Imposing the usual assumptions of an isotropic and homogeneous spacetime, both $g_{\mu\nu}$ and $\tilde{g}_{\mu\nu}$ are given by FRW
metrics with scale factors $a$ and $b=ae^{\overline{\phi}}$, respectively, where $\overline{\phi}$ is the background value of the scalar
field \cite{teves, Skordis2006}. Adopting a spatially flat universe, the modified Friedmann equation in the matter frame reads
\begin{equation}
3H^{2} = 8\pi G_{\rm eff}\left (\overline{\rho}_{\phi} + \overline{\rho}\right ),
\label{eq:17}
\end{equation}
where the physical Hubble parameter is $H=\dot{a}/a^{2}$ and the overdot denotes the derivative with respect to conformal time. Here $\overline{\rho}$
corresponds to the FRW background density of the fluid and the scalar field density takes the form
\begin{equation}
\overline{\rho}_{\phi} = \frac{e^{2\overline{\phi}}}{16\pi G}\left (\overline{\mu}V^{\prime}+V\right ).
\label{eq:18}
\end{equation}
The effective gravitational coupling strength is given by
\begin{equation}
G_{\rm eff} = G e^{-4\overline{\phi}}\left (1+\frac{d\overline{\phi}}{d\log a}\right )^{-2}
\label{eq:19}
\end{equation}
which is generally time-varying through its dependence on the scalar field $\overline{\phi}$. Just as in GR, the energy density $\overline{\rho}$
evolves according to
\begin{equation}
\dot{\overline{\rho}} = -3\frac{\dot{a}}{a}(1+w)\overline{\rho},
\label{eq:20}
\end{equation}
where $w$ is the equation-of-state (EoS) parameter of the fluid.
In case of multiple background fluids, i.e. $\overline{\rho}=\sum_{i}\overline{\rho}_{i}$, the relative densities $\Omega_{i}$ are defined as
\begin{equation}
\Omega_{i} = 8\pi G_{\rm eff}\frac{\overline{\rho}_{i}}{3H^{2}} = \frac{\overline{\rho}_{i}}{\overline{\rho} + \overline{\rho}_{\phi}}.
\label{eq:21}
\end{equation}
The evolution of the scalar field $\phi$ is governed by
\begin{equation}
\ddot{\overline{\phi}} = \dot{\overline{\phi}}\left (\frac{\dot{a}}{a}-\dot{\overline{\phi}}\right )-\frac{1}{U}\left\lbrack 3\overline{\mu}
\frac{\dot{b}}{b}\dot{\overline{\phi}} + 4\pi Ga^{2}e^{-4\overline{\phi}}\left (\overline{\rho}+3\overline{P}\right )\right\rbrack ,
\label{eq:22}
\end{equation}
where $\overline{P}$ is the fluid's background pressure, and the function $U$ is related to the potential $V$,
\begin{equation}
U(\overline{\mu}) = \overline{\mu} + 2\frac{V^{\prime}}{V^{\prime\prime}}.
\label{eq:23}
\end{equation}
In addition, the scalar field obeys the constraint equation
\begin{equation}
\dot{\overline{\phi}}^{2} = \frac{1}{2}a^{2}e^{-2\overline{\phi}}V^{\prime}
\label{eq:24}
\end{equation}
which can be inverted to obtain $\overline{\mu}(a,\overline{\phi},\dot{\overline{\phi}})$. For later use, I also introduce the relation
\begin{equation}
2\frac{\dot{a}}{a}\frac{\dot{b}}{b} - \frac{\ddot{b}}{b} - \overline{\mu}\dot{\overline{\phi}}^{2} = 4\pi Ga^{2}e^{-4\overline{\phi}}
\left (\overline{\rho}+\overline{P}\right )
\label{eq:24b}
\end{equation}
which follows from combining Eq. \eqref{eq:17} with Eq. \eqref{eq:24} and the corresponding Raychaudhuri equation \cite{Skordis2006}. For a
broad class of potentials $V$, one typically finds $\exp{\overline{\phi}}\approx 1$ and $\overline{\rho}_{\phi} \ll 1$ throughout cosmological
history \cite{teves, tevesneutrinocosmo, Bourliot2007}. Therefore, the background evolution is very similar to the standard case of GR, with
only small corrections induced by the scalar field.

\subsection{Tracker solutions of the scalar field}
\label{section32}
For the potentials specified by Eq. \eqref{eq:15}, it has been found that the scalar field exhibits a (stable) tracking behavior and synchronizes
its energy density with the dominant component of the universe \cite{tevesneutrinocosmo, Bourliot2007}. Tracking occurs as $V^{\prime}$ tends to
its zero point where $\overline{\mu}=2\mu_{0}$, and the evolution of the field $\overline{\phi}$ is then approximately given by
\begin{equation}
\overline{\phi} = \overline{\phi}_{0} + \frac{\lvert 1+3w\rvert}{2\beta\mu_{0}\lvert 1-w\lvert - \lvert 1+3w\rvert}\log{a},
\label{eq:25}
\end{equation}
where $\overline{\phi}_{0}$ is an integration constant and $\beta = \pm 1$. The sign depends on the matter fluid's EoS parameter $w$ and Eq.
\eqref{eq:22}. The density $\overline{\rho}_{\phi}$ then exactly scales like that of the fluid, and the relative density parameter $\Omega_{\phi}$
turns approximately into a constant,
\begin{equation}
\Omega_{\phi} = \frac{\left (1+3w\right )^{2}}{6\mu_{0}\left (1-w\right )^{2}}.
\label{eq:26}
\end{equation}
The right-hand side of Eq. \eqref{eq:25} slightly differs from the expression presented in Ref. \cite{Bourliot2007}. In Appendix \ref{appendix1},
I show that Eq. \eqref{eq:25} is indeed the correct result.

During tracking, $\overline{\mu}$ can be expressed as $\overline{\mu} = 2\mu_{0}(1+\epsilon)$
with $0<\epsilon\ll 1$. Using $V^{\prime}(2\mu_{0})=0$ and expanding $V^{\prime}$ to lowest order in $\epsilon$, Eq. \eqref{eq:24} leads to
\begin{equation}
\epsilon = \frac{1}{2}\left (\frac{16\pi l_{B}^{2}}{3\mu_{0}}\frac{e^{2\overline{\phi}}}{a^{2}}\dot{\overline{\phi}}^{2}\right )^{1/n}.
\label{eq:27}
\end{equation}
It turns out that this is the only stage at which the constant $l_{B}$ enters the evolution equations. Taking the time derivative of the above
yields the useful relation
\begin{equation}
\dot{\overline{\phi}}\dot{\epsilon} = \frac{2}{n}\left (\dot{\overline{\phi}}^{2} - \dot{\overline{\phi}}\frac{\dot{a}}{a}
+ \ddot{\overline{\phi}} \right )\epsilon .
\label{eq:27b}
\end{equation}
Stable tracking requires $\epsilon$ to asymptotically decrease to zero, i.e. $\epsilon\rightarrow 0$. Therefore, one has the condition
$\dot{\epsilon} < 0$ which may be used to infer the proper sign of the parameter $\beta$ in Eq. \eqref{eq:25}.

\subsection{Modified Einstein-de Sitter cosmology}
\label{section33}
In what follows, I shall assume a universe entirely made of pressureless matter with perfect tracking of the scalar field, corresponding
to the EdS model in GR. Setting $\overline{P} = w = 0$ fixes $\beta=-1$, and thus the scalar field can be written as
\begin{equation}
\overline{\phi} = \overline{\phi}_{0} - \frac{1}{2\mu_{0}+1}\log{a}.
\label{eq:28}
\end{equation}
To find the proper value of $\beta$, one may either insert Eq. \eqref{eq:25} into Eq. \eqref{eq:22}, or use the argument presented in
Appendix \ref{appendix1}. Since the fluid evolves according to Eq. \eqref{eq:20}, the density takes the form $\overline{\rho}=\overline{\rho}
_{0}a^{-3}$, where $\overline{\rho}_{0}$ is the background density's value today. Exploiting Eq. \eqref{eq:26} allows one to rewrite the
modified Friedmann equation in the matter frame as
\begin{equation}
H^{2} = H_{0}^{2}a^{-3+4/(2\mu_{0}+1)},
\label{eq:29}
\end{equation}
where I have used
\begin{equation}
H_{0}^{2} = e^{-4\overline{\phi}_{0}}\frac{8\pi G\overline{\rho}_{0}}{3}\left (1+\frac{1}{6\mu_{0}-1}\right )\left (1-\frac{1}{2\mu_{0}+1}
\right )^{-2}.
\label{eq:30}
\end{equation}
The deviation of the Hubble expansion from the ordinary EdS case is entirely characterized by the parameter $\mu_{0}$. For several reasons,
$\mu_{0}$ should take a rather large value on the order of $100-1000$ \cite{teves}. Thus this deviation will be small, typically at the
percent level in the range of practical interest.

\section{Metric perturbations in TeVeS}
\label{section4}
\subsection{Preliminaries}
\label{section41}
\subsubsection{Matter-frame perturbations}
Now I will turn to metric perturbations around a spatially flat FRW spacetime in TeVeS. The starting point is the set of linear perturbation
equations derived in Ref. \cite{Skordis2006}. For simplicity, I shall consider only scalar modes and work within the conformal Newtonian
gauge. Perturbations of the metric are then characterized by two scalar potentials $\Psi$ and $\Phi$, and the line element in the matter frame
is given by
\begin{equation}
ds^{2} = a^{2}\left\lbrack -(1+2\Psi)d\tau^{2} + (1-2\Phi )\delta_{ij}dx^{i}dx^{j}\right\rbrack .
\label{eq:31}
\end{equation}
Similarly, one needs to consider perturbations of the other fields: While the fluid perturbation variables are defined in the usual way, i.e.
the density perturbation, for instance, is expressed in terms of the density contrast $\delta$, 
\begin{equation}
\rho = \overline{\rho} + \delta\rho = \overline{\rho}\left (1+\delta \right ),
\label{eq:32}
\end{equation}
the scalar field is perturbed as
\begin{equation}
\phi = \overline{\phi} + \varphi ,
\label{eq:33}
\end{equation}
where $\varphi$ is the scalar field perturbation. Finally, the perturbed vector field is written as
\begin{equation}
A_{\mu} = ae^{-\overline{\phi}}\left (\overline{A}_{\mu} + \alpha_{\mu} \right ) ,
\label{eq:34}
\end{equation}
where $\overline{A}_{\mu} = (1,0,0,0)$ and
\begin{equation}
\alpha_{\mu} = \left (\Psi - \varphi ,\bm{\nabla}\alpha \right ).
\label{eq:35}
\end{equation}
The time component of the vector field perturbation is constrained to be a combination of metric and scalar field perturbations, which is a
consequence of the unit-norm condition in Eq. \eqref{eq:3}. Therefore, one needs to consider only the longitudinal perturbation component $\alpha$.

\subsubsection{Einstein-frame perturbations}
Instead of using Eq. \eqref{eq:31}, one may also express perturbations in the Einstein frame \cite{tevesneutrinocosmo,Skordis2006}. In this
case, metric perturbations are written as
\begin{align}
\tilde{g}_{00} &= -b^{2}e^{-4\overline{\phi}}\left (1+2\tilde{\Psi}\right ),\label{eq:36}\\
\tilde{g}_{0i} &= -b^{2}\partial_{i}\tilde{\zeta},\label{eq:37}\\
\tilde{g}_{ij} &= b^{2}\left (1-2\tilde{\Phi}\right )\delta_{ij},\label{eq:38}
\end{align}
where the Einstein-frame perturbations are given by
\begin{align}
\tilde{\Psi} &= \Psi - \varphi ,\label{eq:39}\\
\tilde{\Phi} &= \Phi - \varphi ,\label{eq:40}\\
\tilde{\zeta} &= \left( e^{-4\overline{\phi}}-1\right )\alpha .\label{eq:41}
\end{align}
To avoid lengthy expressions in the perturbed field equations, it is convenient to work with variables from both frames. Since the equation
governing the evolution of the scalar field perturbation is of second order, it is further helpful to introduce an auxiliary field $\gamma$
which allows one to split the scalar field equation into a system of two first-order equations \cite{Skordis2006}. For the present gauge
choice, the field $\gamma$ is given by
\begin{equation}
\gamma = 2e^{\overline{\phi}}\frac{U}{a}\left (-\dot{\varphi} + \dot{\overline{\phi}}\tilde{\Psi}\right ).
\label{scalareq:2}
\end{equation}

\subsection{Parameterizing the new degrees of freedom}
\label{section43}
To explore how the new degrees of freedom modify cosmological dynamics, it is helpful to define closure relations which may be used
to eliminate perturbations of the scalar and vector field from the evolution equations. Consider a new function $B_{\varphi}$ such
that
\begin{equation}
\log{B_{\varphi}} = \tilde{\Psi} - \varphi .
\label{eq:53}
\end{equation}
The function $B_{\varphi}$ is generally time and scale-dependent, and describes the magnitude of scalar field perturbations
relative to the metric potential $\tilde{\Psi}$. Its definition is not completely arbitrary, but motivated by the algebraic structure
of the perturbation equations. The idea is to try to express the new variables in terms of metric and matter fluid perturbations,
which typically yields equations that are trivially satisfied because of vanishing background terms. The form of these equations
can then serve as a hint for defining suitable relations. Since such a procedure is neither fundamental nor unique, however, one
should just take Eq. \eqref{eq:53} as an educated guess.

Similarly, I define a relation for the variables $\gamma$ and $\tilde{\zeta}$,
\begin{equation}
\log{B_{\gamma}} \propto 3\dot{\tilde{\Phi}} + k^{2}\tilde{\zeta} - \frac{ae^{-\overline{\phi}}}{2U}\gamma,
\label{eq:58}
\end{equation}
where $\mathbf{k}$ is the conformal wave vector, and $k = \lvert\mathbf{k}\rvert$. In what follows, the proportionality constant is
set to unity for simplicity. $B_{\gamma}$ characterizes a combination of both the scalar and the vector field perturbations. Because
of the term $k^{2}\tilde{\zeta}$, the relative contribution due to the vector field perturbation $\alpha$ is scale-dependent and
becomes negligible as $k\rightarrow 0$. Likewise, it will dominate the expression on very small scales within the horizon. Hence
$B_{\gamma}$ may be solely associated with vector field perturbations if one considers the theory on subhorizon scales. The functions
$B_{\varphi}$ and $B_{\gamma}$ are strictly positive by construction, and thus well-defined throughout cosmological history.

Finally, let me also introduce
\begin{equation}
\begin{split}
\tilde{B}_{\varphi} &= \log{B_{\varphi}},\quad \hat{B}_{\varphi}\frac{\dot{a}}{a} = \frac{\dot{B}_{\varphi}}{B_{\varphi}},\\
\tilde{B}_{\gamma}\frac{\dot{a}}{a} &= \log{B_{\gamma}},\quad \hat{B}_{\gamma}\frac{\dot{a}^{2}}{a^{2}} =
\frac{\dot{B}_{\gamma}}{B_{\gamma}},
\end{split}
\label{eq:55}
\end{equation}
which allows one to express the modifications in a more convenient way. In terms of these new functions, the dynamics of GR is
recovered in the limit
\begin{equation}
\tilde{B}_{\varphi}\rightarrow\tilde{\Psi},\quad\tilde{B}_{\gamma}\frac{\dot{a}}{a}\rightarrow 3\dot{\tilde{\Phi}},
\quad\overline{\phi}\rightarrow 0.
\label{grlimit}
\end{equation}

\subsection{Subhorizon scales}
\label{section45}
In the following, I shall assume the previously discussed modified EdS cosmology with perfect tracking of the scalar field. This allows one
to use the corresponding background expressions presented in Sec. \ref{section33} and considerably simplifies the analysis of the modified
equations. Since $\overline{\mu}$ resides close to its minimum in this case, i.e. $\overline{\mu} = 2\mu_{0}(1+\epsilon)$ with $\epsilon\ll
1$, one may further exploit the two first-order expressions
\begin{equation}
\frac{\overline{\mu}}{U} = 1-\frac{2}{n}\epsilon
\label{eq:56}
\end{equation}
and
\begin{equation}
\frac{2\mu_{0}}{U} = 1-\frac{n+2}{n}\epsilon
\label{eq:57}
\end{equation}
which are useful to rewrite terms involving the field $U$.

\subsubsection{Modified potentials}
\label{section45a}
Adopting an EdS universe together with the closure relations presented in Sec. \ref{section43}, one may now write metric perturbations
solely in terms of auxiliary functions and matter variables. The resulting equations are quite lengthy and can be found in Appendix
\ref{appendix3}. As a first application, I consider scales much smaller than the horizon,  i.e. $aH/k\ll 1$. Inserting the logarithmic
approximation for $\overline{\phi}$ specified by Eq. \eqref{eq:28}, I expand the Einstein-frame potentials $\tilde{\Psi}$ and $\tilde{
\Phi}$ in powers of $aH/k$. To second order, this yields
\begin{align}
\tilde{\Psi} &= \mathcal{A}_{0} + \mathcal{A}_{2}\frac{a^{2}H^{2}}{k^{2}} + \mathcal{O}(\epsilon ),\label{persub:5}\\
\tilde{\Phi} &= \mathcal{B}_{0} + \mathcal{B}_{2}\frac{a^{2}H^{2}}{k^{2}} + \mathcal{O}(\epsilon ),\label{persub:6}
\end{align}
where
\begin{align}
\mathcal{A}_{0} &= \frac{\left\lbrack K_{B}-2\left (1-e^{4\overline{\phi}}\right )\right\rbrack\tilde{B}_{\varphi}}{2\left
(1-e^{4\overline{\phi}}
\right )-K_{B}\left (1-2e^{4\overline{\phi}}\right )},\label{persub:7}\\
\mathcal{B}_{0} &= \frac{e^{4\overline{\phi}}K_{B}\tilde{B}_{\varphi}}{2\left (1-e^{4\overline{\phi}}\right )-K_{B}\left
(1-2e^{4\overline{\phi}}
\right )},\label{persub:8}
\end{align}
and $\mathcal{A}_{2}$, $\mathcal{B}_{2}$ are complicated expressions involving $\delta$, $\overline{\phi}$, and the functions in Eq.
\eqref{eq:55}. Compared to GR, where the EdS model gives $\mathcal{A}_{0}=\mathcal{B}_{0}=0$ and $\mathcal{A}_{2}=\mathcal{B}_{2}=
-3\delta/2$, the potentials exhibit a sophisticated dependence on $\overline{\phi}$ which is expected to have a significant impact
on their evolution.

\begin{figure}
\includegraphics[width=\linewidth]{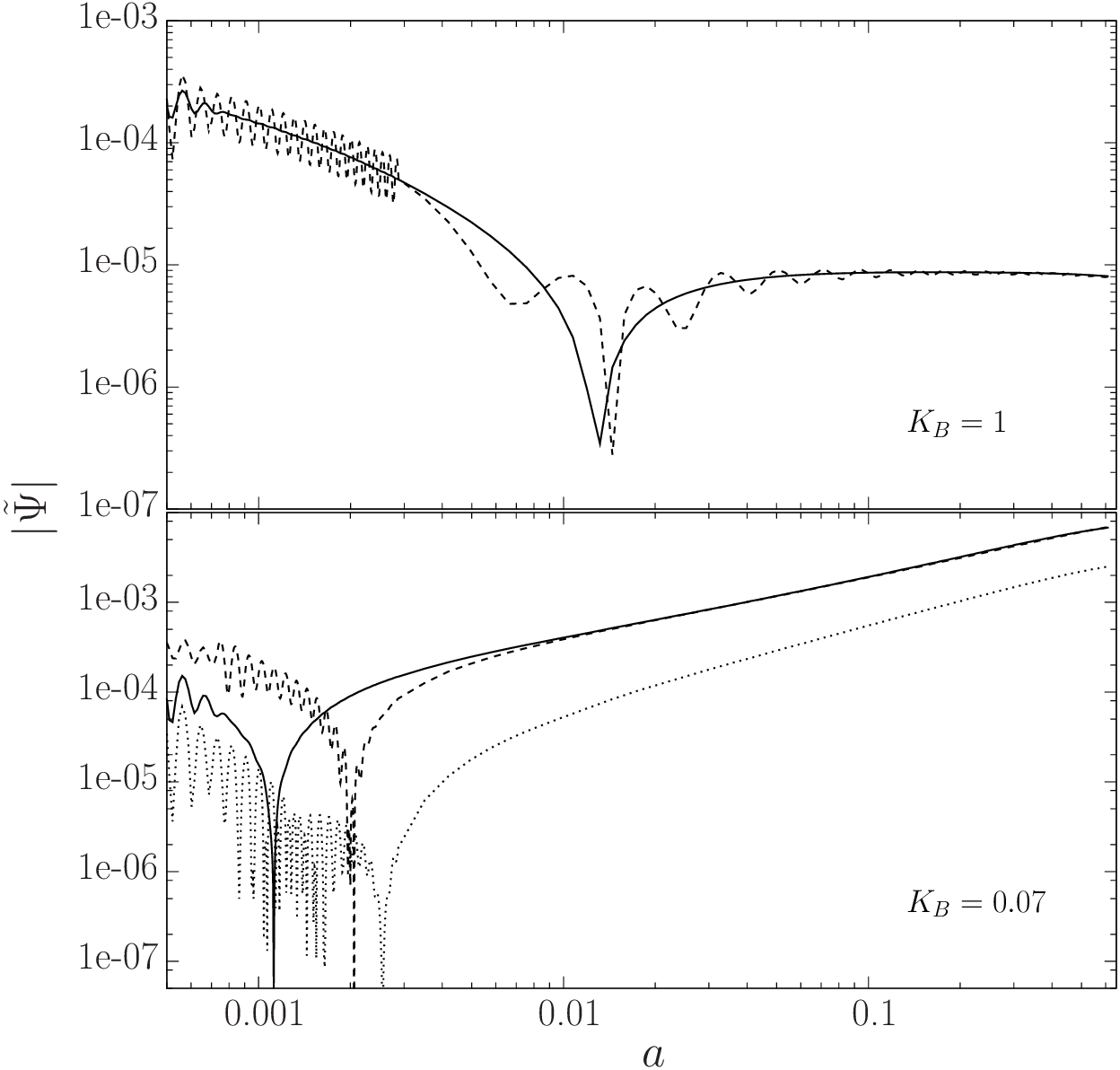}
\caption{Evolution of $\tilde{\Psi}$ for $K_{B}=1$ (top) and $K_{B}=0.07$ (bottom). Assuming $\mu_{0}=250$ and a fixed wavenumber $k=0.5$
Mpc$^{-1}$, the figure shows the full numerical (solid lines) and semi-analytic (dashed lines) results described in the text. The dotted
line indicates the contribution of the zeroth-order term $\mathcal{A}_{0}$.}
\label{fig2}
\end{figure}

Previous studies of TeVeS cosmology found that structure can form more efficiently than in GR and identified a vector instability, which
occurs for small enough values of $K_{B}$, as the key ingredient for enhanced growth \cite{tevesneutrinocosmo, dod1, dod3}. Scalar field
perturbations, on the other hand, were argued to play only a negligible role for structure formation. Here I seek to study this mechanism
at the level of the modified potentials. To this end, I will also consider full numerical solutions which were obtained with the modified
Boltzmann code of Ref. \cite{tevesneutrinocosmo}. These solutions serve as a consistency check of the used assumptions, and allow one to
determine the auxiliary functions $B_{\varphi}$ and $B_{\gamma}$ and to characterize their effect on the potentials. The solver adopts
standard adiabatic initial conditions where perturbations of the TeVeS scalar and vector fields are set to zero. Assuming three neutrinos
with a mass of $2.3$ eV and no dark matter, all calculations in this work use a spatially flat cosmology with density parameters $\Omega_{
m}\approx 0.04$ and $\Omega_{\nu}\approx 0.15$, dimensionless Hubble constant $h=0.74$, and fixed TeVeS parameters $l_{B}=100$ Mpc and
$n=2$.

Let me now focus on $\tilde{\Psi}$ which dictates the dynamics of nonrelativistic matter. Since $\vert\overline{\phi}\vert\ll 1$ and $\mu
_{0}\gg 1$ , Eq. \eqref{persub:5} can be further simplified by taking the limit $\mu_{0}\rightarrow\infty$ \footnote{The value of $\mu_{0}$
still matters for the growth of vector perturbations and enters through the auxiliary functions and $\overline{\phi}$.} and keeping only
leading-order terms with respect to $\overline{\phi}$. In the process, one has to be careful when dealing with expressions involving the
constant $K_{B}$, especially if $K_{B}$ is comparable to $\overline{\phi}$.
\paragraph{$K_{B}\gtrsim 1$.}
In this case, it turns out that
\begin{equation}
\mathcal{A}_{0} \approx \tilde{B}_{\varphi}
\label{persub:9}
\end{equation}
and
\begin{equation}
\mathcal{A}_{2} \approx -2\left (\tilde{B}_{\gamma} + \hat{B}_{\gamma} - 10\tilde{B}_{\varphi} - 4\hat{B}_{\varphi}\right ).
\label{persub:10}
\end{equation}
If the sum of $\tilde{B}_{\gamma}$ and the two time-derivative terms is not too large, the zeroth-order contribution, $\mathcal{A}_{0}$,
will dominate, and $\tilde{\Psi}\approx\tilde{B}_{\varphi}$ during matter domination. The top panel of Fig. \ref{fig2} shows a comparison
between the numerically computed potential and $\tilde{B}_{\varphi}$, assuming $K_{B}=1$, $\mu_{0}=250$, and a wavenumber $k=0.5$ Mpc$^{-1}$.
Indeed, the function $\tilde{B}_{\varphi}$ matches the true potential quite well, which suggests a negligible contribution from vector
field perturbations as long as $K_{B}$ is large enough. Since the scalar field perturbation typically satisfies $\varphi\ll\tilde{\Psi}$,
the approximation $\tilde{\Psi}\approx\tilde{B}_{\varphi}$ also follows independently from the definition of $B_{\varphi}$ and is valid
for any choice of $K_{B}$.

\paragraph{$K_{B}\ll 1$.}
To leading order, one finds
\begin{equation}
\mathcal{A}_{0} \approx \frac{1 - 2\tilde{K}^{-1}}{1 + 2\tilde{K}^{-1}}\tilde{B}_{\varphi}
\label{persub:11}
\end{equation}
and
\begin{equation}
\begin{split}
&\mathcal{A}_{2} \approx 4\frac{5 - 41\tilde{K}^{-1} + 56\tilde{K}^{-2} - 32\tilde{K}^{-3}}{\left (1 + 2\tilde{K}^{-1}\right )^{2}}
\tilde{B}_{\varphi}\\
&+ \tilde{K}^{-1}\frac{4\left (\tilde{B}_{\gamma} + \hat{B}_{\gamma} - 5\hat{B}_{\varphi}\right ) - 6\delta}{1 + 2\tilde{K}^{-1}}
- 2\frac{\tilde{B}_{\gamma} +\hat{B}_{\gamma} - 4\hat{B}_{\varphi}}{1 + 2\tilde{K}^{-1}},
\end{split}
\label{persub:12}
\end{equation}
where $\tilde{K}^{-1}K_{B} = 1-e^{4\overline{\phi}}$. There are two main differences with respect to the previous case. First, $\delta$
explicitly appears in the equation, sourcing the potential directly and not only through the functions $B_{\varphi}$ and $B_{\gamma}$.
Secondly, the terms $\mathcal{A}_{0}$ and $\mathcal{A}_{2}$ now strongly depend on the ratio $\tilde{K}^{-1}$. Calculations for the
background yield $\overline{\phi}<0$ and $\tilde{K}^{-1}>0$, resulting in a suppression of $\mathcal{A}_{0}$ relative to higher-order
contributions. Hence, the second-order term, specified by $\mathcal{A}_{2}$, can be expected to be important. The bottom panel of Fig.
\ref{fig2} shows $\tilde{\Psi}$ together with the semi-analytic approximation using Eqs. \eqref{persub:11} and \eqref{persub:12} for
$K_{B}=0.07$, $\mu_{0}=250$, and $k=0.5$ Mpc$^{-1}$. The full numerical result is again matched very well, but this time the contribution of $\mathcal{
A}_{0}$ is much smaller and negative, growing up to a fraction of roughly 30\% near the end of matter domination. Considering the term
$\mathcal{A}_{2}$, only expressions related to $B_{\gamma}$ and $\delta$ turn out as relevant sources.

The above demonstrates that the algebraic mechanism triggered by small values of the vector coupling, $K_{B}\sim\vert 4\overline{\phi}
\vert$, is an important element in generating enhanced growth. Over the range $0.07<K_{B}<1$, the full numerical analysis shows that
the ratio $\tilde{B}_{\gamma}/\tilde{B}_{\varphi}$ changes only by a factor of less than 3--4 (and similarly for $\hat{B}_{\gamma}/
\hat{B}_{\varphi}$), not enough to explain why the terms involving $B_{\varphi}$ and its time derivative are suppressed. Neglecting
the new contributions and changes in the above equations does not lead to an augmentation of growth. In accordance with Ref. \cite{dod1},
this points toward the implementation of the TeVeS vector field as the key ingredient for boosting structure formation. Once enhanced
growth occurs, the potential is found to be sourced by $B_{\varphi}$, $B_{\gamma}$, and $\delta$ at comparable levels. As the evolution
equations of the potentials, the scalar and vector fields, and the density contrast are coupled to each other, however, this complex
interplay is hardly surprising.

\begin{figure}
\includegraphics[width=\linewidth]{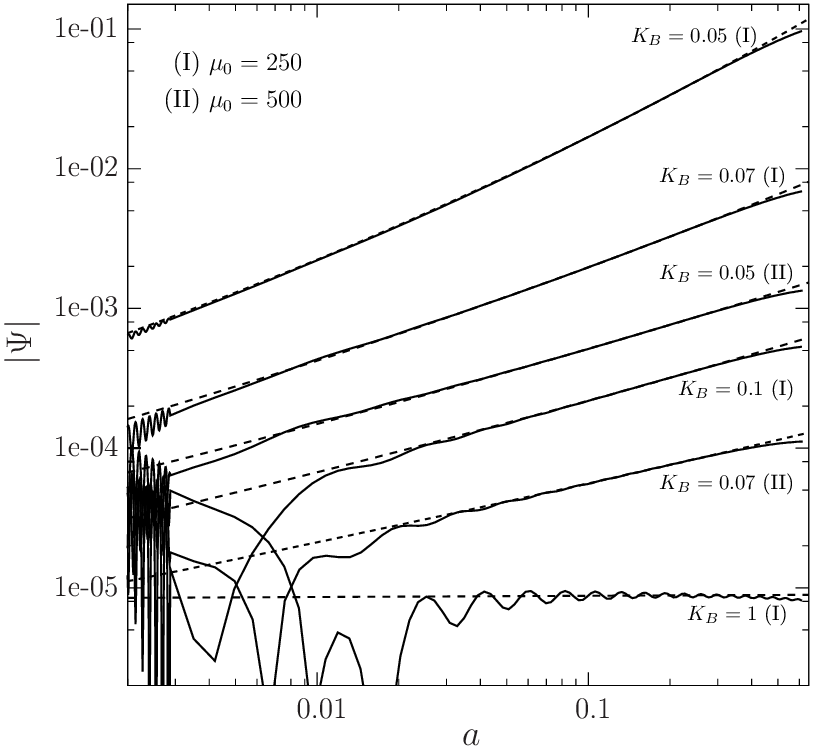}
\caption{Evolution of $\Psi$ for $k=0.5$ Mpc$^{-1}$ and different combinations of $K_{B}$ and $\mu_{0}$. The figure illustrates
the numerical result (solid lines) together with the analytic approximation (dashed lines) valid in the late matter era.}
\label{fig3}
\end{figure}

How do the potentials $\Psi$ and $\tilde{\Psi}$ evolve with cosmic time? To find an answer, I start from the observation that $\tilde{
B}_{\varphi}\sim\hat{B}_{\varphi}$ in the late matter era (and similarly for $\tilde{B}_{\gamma}$ and $\hat{B}_{\gamma}$). A closer
inspection suggests the relation
\begin{equation}
\tilde{B}_{\varphi}^{\prime} \propto a^{p-1}\tilde{B}_{\varphi},
\label{persub:13}
\end{equation}
where the prime denotes the derivative with respect to the scale factor $a$, and the constant $p$ depends on the parameters of the
theory. Remarkably, one finds
\begin{equation}
p^{-1} = \mu_{0}K_{B}
\label{persub:13b}
\end{equation}
which, using the tracking solution Eq. \eqref{eq:28} and $\vert\overline{\phi}\vert\ll 1$, can be related to the
background scalar field through
\begin{equation}
a^{p} \propto e^{-2\overline{\phi}/K_{B}}.
\label{persub:13c}
\end{equation}
Integrating by separation of variables then leads to
\begin{equation}
\log{\tilde{B}_{\varphi}} = c_{1}a^{p} + c_{2},
\label{persub:14}
\end{equation}
where $c_{1}$ is fixed by the proportionality constant and $c_{2}$ is a constant of integration. Since $\varphi\ll\tilde{\Psi}$,
one has $\Psi\approx\tilde{\Psi}\approx\tilde{B}_{\varphi}$, and the potentials approximately satisfy Eq. \eqref{persub:14}.
Figure \ref{fig3} illustrates the evolution of $\Psi$ together with the analytic approximation, assuming $k=0.5$ Mpc$^{-1}$ and
different combinations of $K_{B}$ and $\mu_{0}$. The simple models agree well with the numerical results for $a\sim 0.01$ and
later. In the case of $\mu_{0}K_{B}=250$, the additional growth is strongly suppressed and $\Psi$ basically approaches a constant.
The result is found to be independent of scale as long as $aH/k\ll 1$ is satisfied.
Given the complexity of TeVeS, the appearance of such an effective description is astonishing. Since the growth of perturbations
is mainly determined by the ratio $\overline{\phi}/K_{B}$, one may expect this behavior also for other parameter choices.
Exploring its origin is beyond the scope of this paper, but might be addressed in future work.

\begin{figure}
\includegraphics[width=\linewidth]{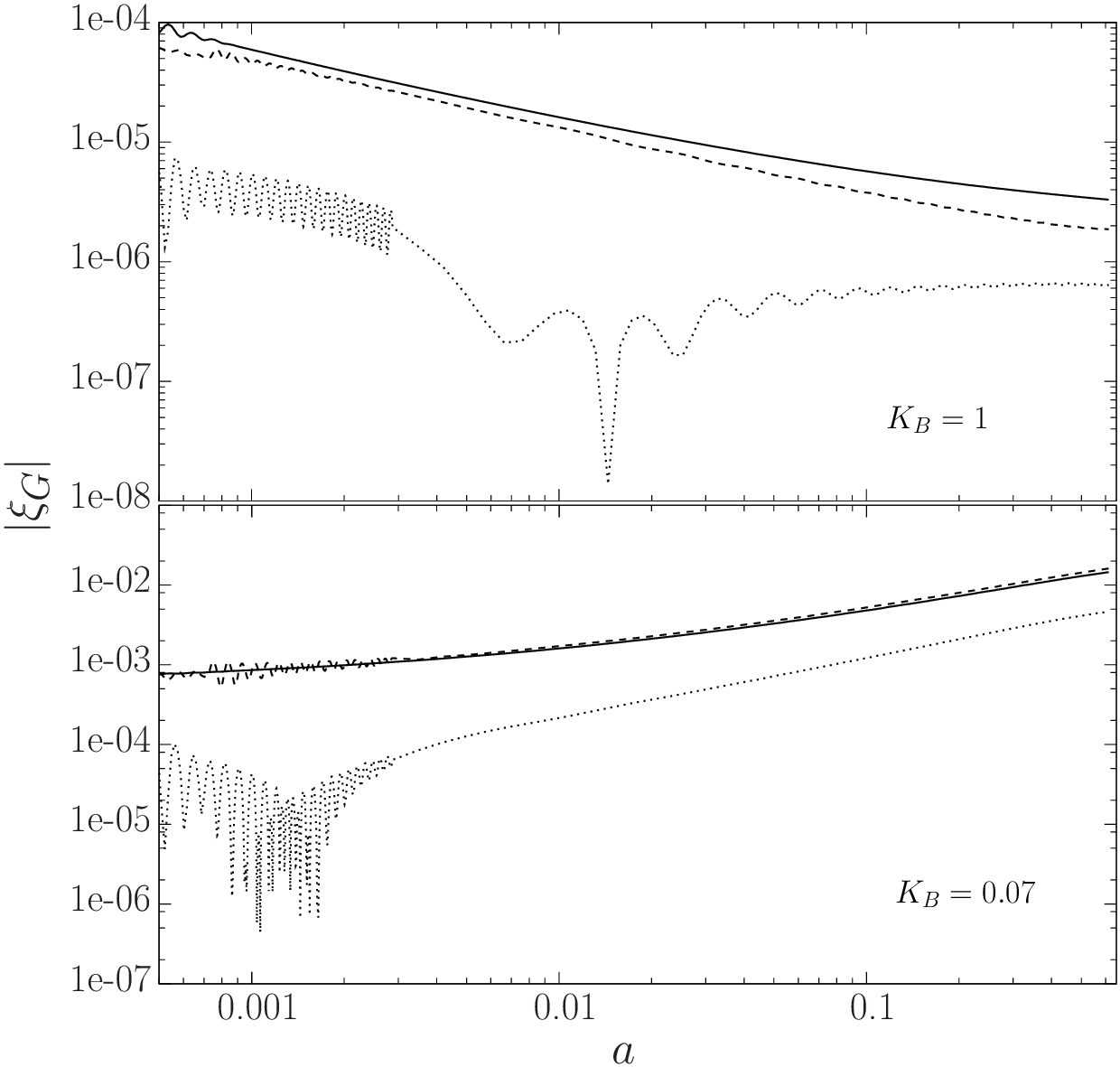}
\caption{Same as Fig. \ref{fig2}, but now for the slip $\xi_{G}$. The dotted lines indicate the contribution of the
zeroth-order term $\mathcal{A}_{0}-\mathcal{B}_{0}$.}
\label{fig4}
\end{figure}

\subsubsection{Gravitational slip}
\label{section45b}
Next, I want to address how the TeVeS growth mechanism manifests itself in a difference between the matter-frame potentials
$\Psi$ and $\Phi$. Introducing the gravitational slip as
\begin{equation}
\xi_{G} = \Psi - \Phi = \tilde{\Psi} - \tilde{\Phi},
\label{slipeq:1}
\end{equation}
I will, just as before, consider the limit $\mu_{0}\rightarrow\infty$ and keep only leading-order terms with respect to
$\overline{\phi}$. All comparisons to full numerical results assume $\mu_{0}=250$ and $k=0.5$ Mpc$^{-1}$.

\paragraph{$K_{B}\gtrsim 1$.}
Collecting the relevant terms yields
\begin{equation}
\mathcal{A}_{0} - \mathcal{B}_{0} \approx \left (e^{4\overline{\phi}} - 1\right )\tilde{B}_{\varphi}
\label{slipeq:2}
\end{equation}
and
\begin{equation}
\mathcal{A}_{2} - \mathcal{B}_{2} \approx \tilde{B}_{\gamma} + 4\tilde{B}_{\varphi} + \hat{B}_{\varphi}.
\label{slipeq:3}
\end{equation}
Unlike the case of $\tilde{\Psi}$, a factor of $e^{4\overline{\phi}} - 1$ suppresses the term $\mathcal{A}_{0} - \mathcal{B}_{0}$,
and hence it should become comparable to higher-order contributions. The top panel of Fig. \ref{fig4} compares $\xi_{G}$ to its
semi-analytic approximation using Eqs. \eqref{slipeq:2} and \eqref{slipeq:3} for $K_{B}=1$. Qualitatively, the approximation is in
agreement with the full numerical result. As expected, the term $\mathcal{A}_{0} - \mathcal{B}_{0}$ is not dominant and increases
to a maximum relative contribution of around 25\% at the end of the matter era. The numerical results show that most of $\xi_{G}$
is due to the second-order expression specified by $\mathcal{A}_{2} - \mathcal{B}_{2} \approx\tilde{B}_{\gamma}$ which, according
to Eq. \eqref{eq:58}, can be associated with the vector field perturbation $\tilde{\zeta}$.

\paragraph{$K_{B}\ll 1$.}
A straightforward calculation leads to
\begin{equation}
\mathcal{A}_{0} - \mathcal{B}_{0} \approx -\frac{2\tilde{K}^{-1}}{1 + 2\tilde{K}^{-1}}\tilde{B}_{\varphi}
\label{slipeq:4}
\end{equation}
and
\begin{equation}
\begin{split}
&\mathcal{A}_{2} - \mathcal{B}_{2} = 2\frac{2 - 29\tilde{K}^{-1} + 80\tilde{K}^{-2} - 56\tilde{K}^{-3}}{\left (1 + 2\tilde{K}^{-1}
\right )^{2}}\tilde{B}_{\varphi}\\
&+ \tilde{K}^{-1}\frac{4\left (\hat{B}_{\gamma} - 4\hat{B}_{\varphi}\right ) + 3\left (2\tilde{B}_{\gamma}
- \delta\right )}{1 + 2\tilde{K}^{-1}} + \frac{\tilde{B}_{\gamma} +\hat{B}_{\varphi}}{1 + 2\tilde{K}^{-1}}.
\end{split}
\label{slipeq:5}
\end{equation}
For sufficiently small values of $K_{B}$, the effect on $\xi_{G}$ is very similar to the case of $\tilde{\Psi}$, but now both the
first and second-order terms increase to nearly the same degree. The bottom panel of Fig. \ref{fig4} shows $\xi_{G}$ together with
the semi-analytic approximation using Eqs. \eqref{slipeq:4} and \eqref{slipeq:5} for $K_{B}=0.07$. Again, the full numerical result
is well reproduced, and the second-order term, determined by $\mathcal{A}_{2}-\mathcal{B}_{2}$, is dominated by the terms involving
$B_{\gamma}$, its time derivative, and $\delta$. The relative contribution of $\mathcal{A}_{0}-\mathcal{B}_{0}$ is almost identical
to the previous case $K_{B}=1$. Apart from serving as a consistency check and validation of the used assumptions, the above results
confirm that the mechanisms responsible for boosting the potential $\Psi$ and for generating a considerable slip $\xi_{G}$ are,
indeed, the very same.

\begin{figure}
\includegraphics[width=\linewidth]{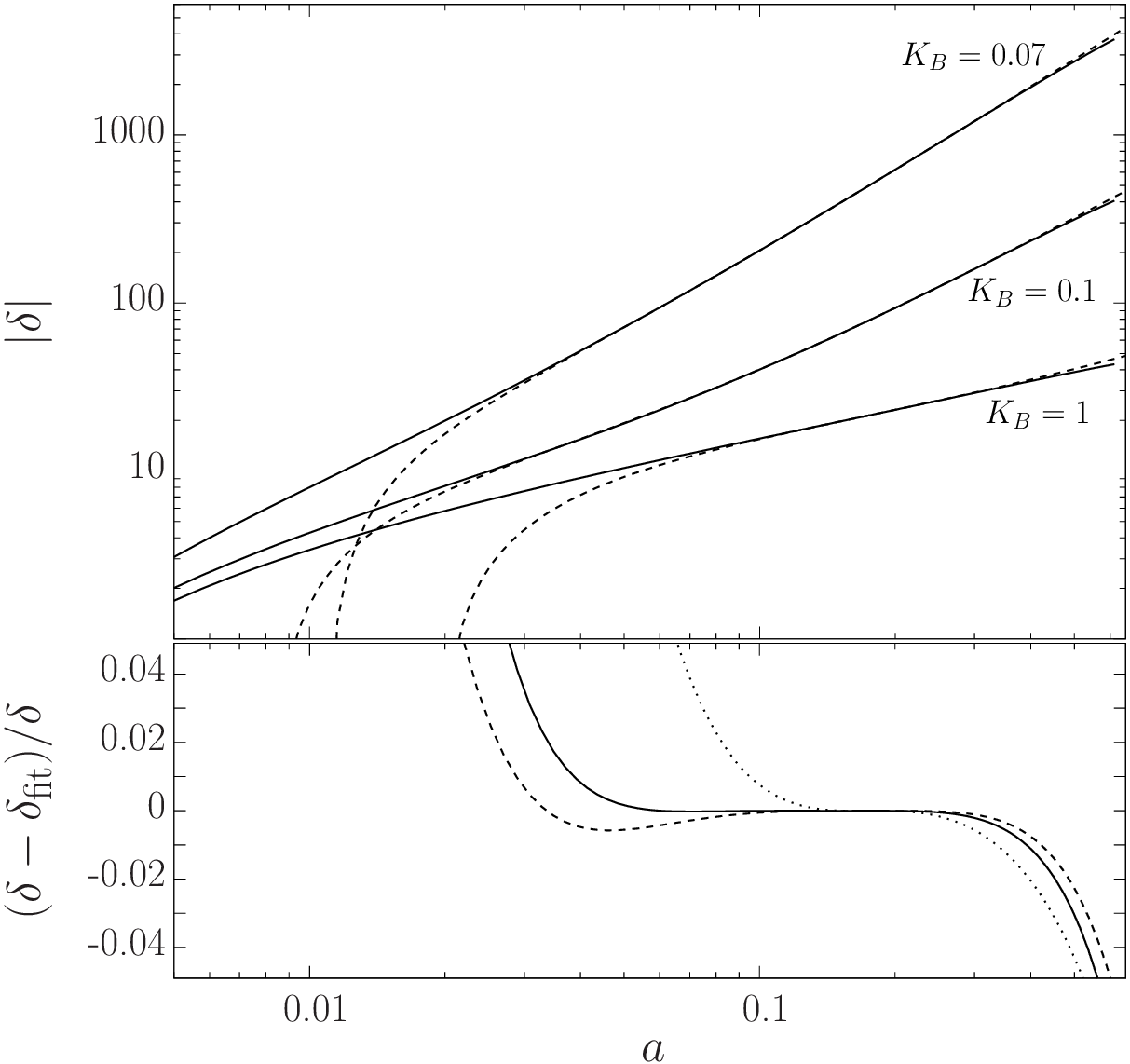}
\caption{Growth of density perturbations for different values of $K_{B}$, $\mu_{0}=250$, and $k=0.5$ Mpc$^{-1}$. The top panel
shows the evolution of $\delta$ (solid lines) together with analytic models based on approximate solutions to the modified growth
equation (dashed lines; see text). The bottom panel depicts the models' relative accuracy for $K_{B}=0.07$ (solid
line), $0.1$ (dashed line), and $1$ (dotted line).}
\label{fig5}
\end{figure}

\subsubsection{Growth of density perturbations}
\label{section45c}
To conclude the section on subhorizon scales, consider the growth of density perturbations during the matter-dominated phase.
Since only the gravitational part is modified, one may follow the usual derivation of the growth equation and obtains
\begin{equation}
a^{2}\delta^{\prime\prime} + \frac{a}{2}\left (3+\frac{4}{2\mu_{0}+1}\right )\delta^{\prime}
+ \left (\frac{k}{aH}\right )^{2}\Psi = 0.
\label{per2eq:1}
\end{equation}
If $p=(\mu_{0}K_{B})^{-1}$ is chosen small enough, the analytic expression of $\Psi$ given by Eq. \eqref{persub:14} can be
approximated by its first-order expansion. Using $aH\propto a^{-1/2}$ then yields a relation of the form
\begin{equation}
\left (\frac{k}{aH}\right )^{2}\Psi \propto a^{q},\quad q - 1 \propto p.
\label{per2eq:2}
\end{equation}
Substituting this into Eq. \eqref{per2eq:1} and remembering $\mu_{0}\gg 1$, the solution for $\delta$ can be written as the
sum of a power-law, i.e. $\delta\propto a^{q}$, and a linear combination of solutions to the homogeneous system,
\begin{equation}
\delta_{1}\propto a^{-1/2},\quad \delta_{2}={\rm const}.
\label{per2eq:3}
\end{equation}
The approximation is expected to hold even slightly beyond the validity of the adopted first-order expansion since $\log{\Psi}$
effectively grows proportionally to $\log{a}$ over the time range of interest in cases where $p\leq 0.05$.

The top panel of Fig. \ref{fig5} illustrates the evolution of $\delta$ together with these analytic models (dashed lines) for
$k=0.5$ Mpc$^{-1}$, $\mu_{0}=250$, and different choices of $K_{B}\geq 0.07$. Starting from $a\sim 0.02$, the numerical results
nicely tend toward the analytic solutions which were matched near the end of matter domination at $a=0.2$. Their relative deviation
is indicated in the figure's bottom panel, and is in accordance with Sec. \ref{section45a}, where similar domains of validity
have been found. If $p\ll 1$, the constant $q$ approaches unity and one obtains the usual GR EdS growing mode $\delta\propto a$,
which is the case for $K_{B}=1$. The other cases exhibit enhanced growth and are characterized by power-law solutions with $q>1$.
Again, it is remarkable that the complex dynamics of TeVeS allows a surprisingly simple, effective description of perturbation
growth for a wide range of parameters.

For $\mu_{0}K_{B}\sim 10$ and smaller, the simple power-law approximation will break down and one must resort to Eq. \eqref{persub:14}
when solving the growth equation, leading to analytic solutions for $\delta$ which involve exponential integrals. As is seen from the
topmost line in Fig. \ref{fig3}, $\Psi$ grows faster than a power law in these cases, and one may expect a similar behavior for $\delta$.

\subsection{Superhorizon scales}
\label{section44}
On scales much larger than the horizon, terms proportional to $k^{2}$ may be safely neglected in the perturbation equations,
i.e. $k\rightarrow 0$. Considering the Hamiltonian constraint and the evolution equation for $\delta$, one obtains
\begin{equation}
-3\frac{\dot{b}}{b}\left (\dot{\tilde{\Phi}} + \frac{\dot{b}}{b}\tilde{\Psi} \right )+ \frac{ae^{-\overline{\phi}}}{2}
\dot{\overline{\phi}}\gamma = 4\pi Ga^{2}e^{-4\overline{\phi}}\overline{\rho}\left (\delta - 2\varphi \right )
\label{supereq:1}
\end{equation}
and
\begin{equation}
\dot{\delta} = 3\dot{\Phi} = 3\left (\dot{\tilde{\Phi}}+\dot{\varphi}\right ).
\label{supereq:2}
\end{equation}
As usual, the above implies that the combination $\delta - 3\Phi$ is conserved over time.

Using the above equations together with the closure relations from Sec. \ref{section43}, it is possible to arrive at an equation
governing the evolution of $\tilde{\Phi}$. The corresponding derivation is sketched in Appendix \ref{appendix31}, and the final
expression takes the form
\begin{equation}
\tilde{\Phi}^{\prime\prime} + \frac{13}{10a}\tilde{\Phi}^{\prime} = \frac{2}{15a^{2}}\left (\tilde{B}_{\gamma} + 2\hat{B}_{\gamma}
- \frac{3}{2}\hat{B}_{\varphi}\right ),
\label{supereq:3}
\end{equation}
where it is again assumed that $\mu_{0}\gg 1$, and also $\tilde{\Phi}\approx\Phi$ since $\varphi\ll\tilde{\Phi}$ is typically
satisfied. There is no explicit dependence on $K_{B}$ which can only enter through $B_{\varphi}$ and $B_{\gamma}$. Formally,
the solution to Eq. \eqref{supereq:3} can be expressed in terms of the auxiliary functions on the right-hand side which are
unknown a priori. The numerical analysis, however, suggests that the right-hand terms nearly cancel each other,
\begin{equation}
\tilde{B}_{\gamma} + 2\hat{B}_{\gamma} - \frac{3}{2}\hat{B}_{\varphi} \approx 0.
\label{supereq:4}
\end{equation}
Considering the resulting homogeneous equation then leads to the solutions
\begin{equation}
\Phi_{1} \propto a^{-3/10},\quad \Phi_{2} = {\rm const},
\label{supereq:5}
\end{equation}
and one is left with the usual result that $\Phi_{2}$ and $\delta_{2}$ are frozen for modes which have not yet entered the
horizon. Compared to the situation in GR, where $\Phi_{1} \propto a^{-5/2}$, the decaying mode evolves much more slowly and
may not die out fast enough to be discarded in the late matter era.

\begin{figure}
\includegraphics[width=\linewidth]{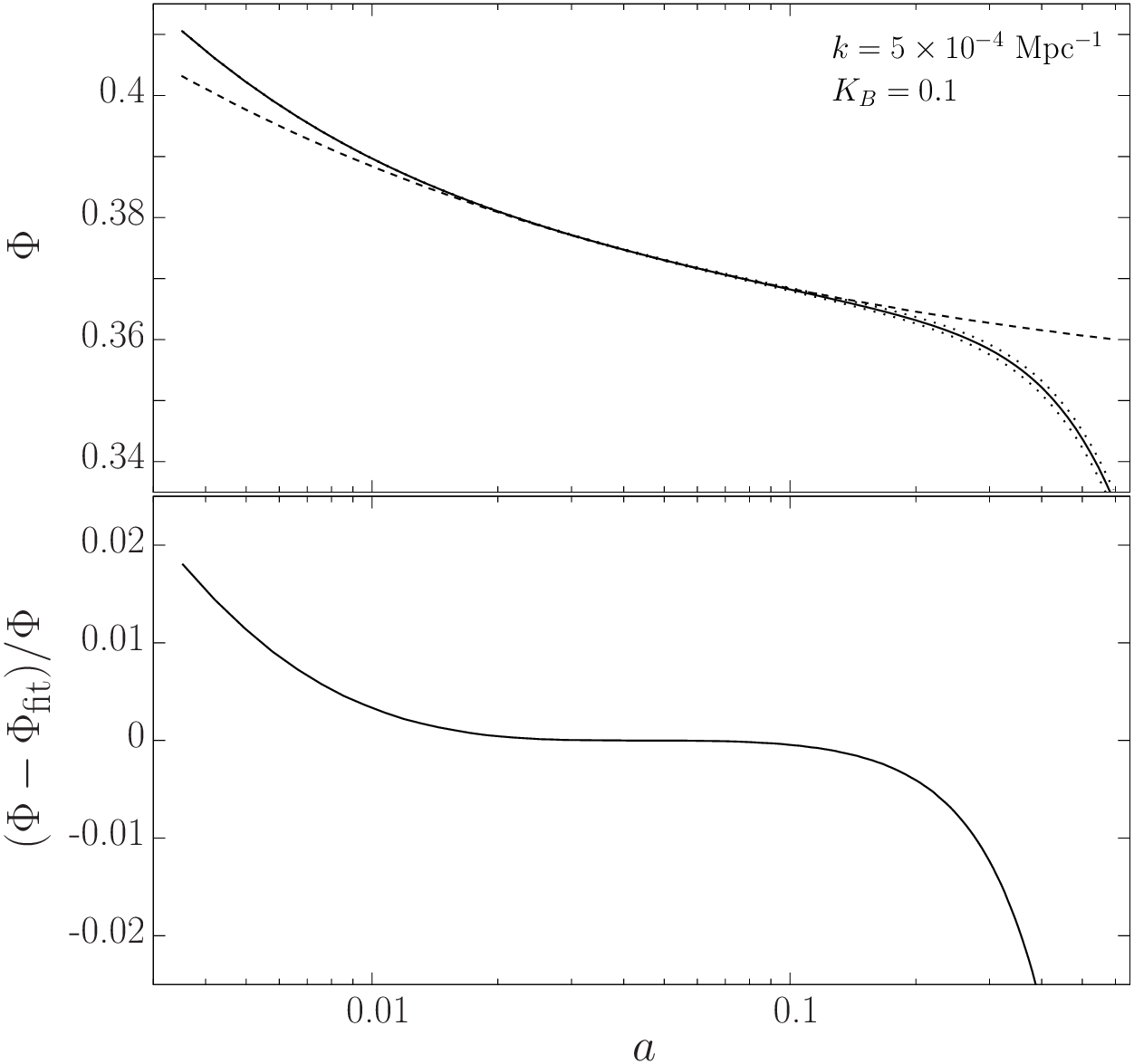}
\caption{Evolution of $\Phi$ for $k=5\times 10^{-4}$ Mpc$^{-1}$, $\mu_{0}=250$, and $K_{B}=0.1$. The top panel shows the numerical
result (solid line) and the analytic model described in the text (dashed line). The dotted lines indicate how $\Phi$ changes within
the range $0.07<K_{B}<1$. The bottom panel depicts the relative accuracy of the analytic model.}
\label{fig6}
\end{figure}

The top panel of Fig. \ref{fig6} shows the evolution of the potential $\Phi$ together with the analytic model (matched near
$a=0.05$) based on Eq. \eqref{supereq:5} for $k=5\times 10^{-4}$ Mpc$^{-1}$, $\mu_{0}=250$, and $K_{B}=0.1$. The numerical
result is basically independent of $K_{B}$, as indicated by the dotted lines which quantify the change of $\Phi$ over the
range $0.07<K_{B}<1$. The analytic model is very accurate, with a relative error less than approximately 2--3\% throughout
matter domination. As already suspected, the decaying mode is important even at late times around $a\sim 0.1$. Hence, $\Phi$
and $\delta$ are no longer conserved, but characterized by a rather slow decay outside the horizon. The insensitivity to $K_{B}$
on these large scales can also be understood from Eq. \eqref{eq:58} which suggests that the impact of vector field perturbations
becomes negligible as $k\rightarrow 0$.

\section{Conclusions}
\label{section5}
In this work, I have revisited the evolution of cosmological perturbations in Bekenstein's TeVeS theory. Considering only scalar
modes in the conformal Newtonian gauge, I have introduced two auxiliary functions, $B_{\varphi}$ and $B_{\gamma}$, which allow
one to express perturbations of the TeVeS scalar and vector fields in a way suitable for studying modifications due to these new
degrees of freedom at the level of the metric potentials.

Assuming a universe in the matter-dominated phase, I have examined the theory's behavior on scales well inside and outside the
horizon, respectively. Deriving approximate expressions for the potentials on subhorizon scales, I have adopted a semi-analytic
approach to identify and describe the mechanism responsible for the superlinear growth of density perturbations. An important
element of this growth mechanism is the ratio between $\overline{\phi}$, the background scalar field, and the coupling constant
$K_{B}$, triggering significant changes in the potentials if $K_{B}\lesssim 4\vert\overline{\phi}\vert$. The analysis confirms
the implementation of the TeVeS vector field as the key to enhanced structure formation, in agreement with the result of Ref.
\cite{dod1}. The evolution of $\delta$, the density contrast, and the potential $\Psi$ during the matter era has been explored
semi-analytically and is well approximated in terms of simple power-law solutions for a wide range of parameters. This is striking
in view of the theory's complexity and does warrant further investigation. In the limit of superhorizon scales, vector field
perturbations effectively decouple from the equations. Unlike the situation in GR, the potential $\Phi$ and $\delta$ are not
conserved outside the horizon, but characterized by a slowly decaying mode which remains important even at late times.

The approach taken here is not restricted to the framework of TeVeS, and might be useful to study the properties of other
modified gravity theories with additional degrees of freedom.

\begin{acknowledgments}
The author likes to thank Jacob D. Bekenstein, Cosimo Fedeli, David F. Mota, Stephen E. Rafter, and the anonymous referees for useful discussions
and comments which significantly improved the manuscript. The author is particularly grateful to Constantinos Skordis for help with TeVeS
perturbations and his modified Boltzmann code. This work was partially supported by a fellowship from the Lady Davis Foundation and by the grant
Spin(e) ANR-13-BS05-0005 of the French National Research Agency. The author acknowledges support through a fellowship from the Minerva Foundation.
\end{acknowledgments}

\appendix

\section{Scalar field evolution during tracking}
\label{appendix1}
In the following, I will assume the generalized potential defined in Eq. \eqref{eq:15} and adopt the notation and definitions used in Ref.
\cite{Bourliot2007}. There it has been found that the scalar field evolves during tracking as
\begin{equation}
\overline{\phi} = \overline{\phi}_{0} + {\phi}_{1}\log{a},
\label{eq:a1}
\end{equation}
where
\begin{equation}
\phi_{1} \equiv \frac{d\overline{\phi}}{d\log a}
\label{eq:a2}
\end{equation}
is approximately constant. Following the derivation presented in Ref. \cite{Bourliot2007}, one shows that
\begin{equation}
\frac{\phi_{1}}{1+\phi_{1}} = \frac{\beta}{2\mu_{0}}\sqrt{\left (\frac{1+3w}{1-w} \right )^{2}},
\label{eq:a3}
\end{equation}
where $\beta =\pm 1$ denotes the sign of the scalar field's time derivative, i.e.
\begin{equation}
\beta \equiv \sgn{\dot{\overline{\phi}}}.
\label{eq:a4}
\end{equation}
To see that the sign in Eq. \eqref{eq:a3} is chosen appropriately, one uses Eq. \eqref{eq:a1} and finds that
\begin{equation}
\beta = \sgn\left ({\phi}_{1}\frac{\dot{a}}{a}\right ) = \sgn{{\phi}_{1}} = \sgn{\beta},
\label{eq:a5}
\end{equation}
where I have assumed that $\lvert\phi_{1}\rvert \ll 1$ for the last equality. Note that this is justified because of the requirement
$\mu_{0} \gg 1$ for viable cosmological models. When evaluating the square root in Eq. \eqref{eq:a3}, one needs to take into account that
the argument's sign depends on the actual choice of $w$. Therefore, one has
\begin{equation}
\sqrt{\left (\frac{1+3w}{1-w} \right )^{2}} = \frac{\lvert 1+3w\rvert}{\lvert 1-w\rvert},
\label{eq:a6}
\end{equation}
which eventually gives the result in Eq. \eqref{eq:25}. During tracking, the field $\overline{\mu}$ (see Sec. \ref{section32}) evolves
as $\overline{\mu} = 2\mu_{0}(1+\epsilon)$, where
\begin{equation}
\log{\epsilon}\propto -\frac{2\phi_{1}+3(1+w)}{n}\log{a},
\label{eq:a7}
\end{equation}
and thus $2\phi_{1}+3(1+w) > 0$ emerges as a condition for stable tracking. For a universe dominated by a cosmological constant $\Lambda$,
one has $w=-1$ and therefore $\beta=1$. Since the time derivative of $\overline{\phi}$ changes its sign when passing from the matter to
the $\Lambda$ era (resulting in $\overline{\rho}_{\phi}$ momentarily going to zero) \cite{tevesneutrinocosmo}, it follows that $\beta = -1$
during matter domination. This result is in accordance with previous work \cite{teves, dod1} where it has been shown that $\overline{\phi}$
decreases with time during the matter era.

\section{Parameterized perturbation equations}
\label{appendix3}
In what follows, I will adopt the linear TeVeS perturbation equations for scalar modes in the conformal Newtonian gauge \cite{Skordis2006}.
Further, I will assume a spatially flat spacetime and a universe filled with pressureless matter only, corresponding to the modified EdS
cosmology introduced in Sec. \ref{section33}. In this case, the fluid's pressure components may be neglected ($w=C_{s}=\Sigma =0$, where
$C_{s}$ and $\Sigma$ are the fluid's sound speed and shear, respectively) and the background density evolves as $\overline{\rho}\propto
a^{-3}$. As usual, the equations are expressed in Fourier space using the conformal wave vector $\mathbf{k}$ in accordance with the
coordinate system specified in Sec. \ref{section41}.

\subsection{General case}
\label{appendix32}
The goal is to express the metric potentials in terms of matter fluid variables, using the closure relations from Sec. \ref{section43}.
Since there remain three fields, i.e. the two metric potentials and the vector field perturbation $\alpha$, this requires finding three
linearly independent equations. For the first equation, I eliminate $\dot{\alpha}$ between the corresponding vector field and propagation
equations, which leads to
\begin{equation}
\tilde{\Phi} + \left (1-e^{4\overline{\phi}}\right )E - e^{4\overline{\phi}}\tilde{\Psi} - 4\dot{\overline{\phi}}\alpha + e^{4\overline{\phi}}
\left (\frac{\dot{a}}{a} + 5\dot{\overline{\phi}}\right )\tilde{\zeta} = 0.
\label{persub:2}
\end{equation}
Differentiating the above and substituting all remaining time derivatives by a suitable combination of the perturbation equations eventually
gives
\begin{widetext}
\begin{equation}
\begin{split}
&\phantom{+e}e^{4\overline{\phi}}k^{2}\tilde{\zeta} + \left\lbrack\frac{\left (1-e^{4\overline{\phi}}\right )^{2}}{K_{B}}
\left (\frac{\overline{\mu}\dot{\overline{\phi}}^{2}}{1-e^{4\overline{\phi}}}+8\pi Ga^{2}e^{-4\overline{\phi}}\overline{\rho}\right )-8\left
(1+2\frac{1+e^{4\overline{\phi}}}{1-e^{4\overline{\phi}}}\right )\dot{\overline{\phi}}^{2}-4\ddot{\overline{\phi}}\right\rbrack\alpha+4\dot{
\overline{\phi}}\left (-1+\frac{1+e^{4\overline{\phi}}}{1-e^{4\overline{\phi}}}\right )\tilde{\Phi}\\
&+e^{4\overline{\phi}}\left\lbrack 4\frac{1+e^{4\overline{\phi}}}{1-e^{4\overline{\phi}}}\dot{\overline{\phi}}\left (\frac{\dot{a}}{a}+5\dot{
\overline{\phi}}\right )+\frac{\ddot{a}}{a}-2\frac{\dot{a}^{2}}{a^{2}}-4\dot{\overline{\phi}}\frac{\dot{a}}{a} +5\dot{\overline{\phi}}^{2}+5
\ddot{\overline{\phi}}\right\rbrack\tilde{\zeta}+ 4\pi Ga^{2}e^{-4\overline{\phi}}\overline{\rho}\left\lbrack 1+3e^{4\overline{\phi}}-2\frac{
\left (1-e^{4\overline{\phi}}\right )^{2}}{K_{B}}\right\rbrack\theta - e^{4\overline{\phi}}\frac{\dot{a}}{a}\hat{B}_{\varphi}\\
&+\left\lbrack\left (1+3e^{4\overline{\phi}}-\frac{1-e^{4\overline{\phi}}}{K_{B}}\right )\overline{\mu}\dot{\overline{\phi}}-e^{4\overline{
\phi}}\left (4\frac{\dot{b}}{b}+5\dot{\overline{\phi}}+4\frac{1+e^{4\overline{\phi}}}{1-e^{4\overline{\phi}}}\dot{\overline{\phi}}\right )
\right\rbrack\tilde{\Psi}-\overline{\mu}\dot{\overline{\phi}}\left (1+3e^{4\overline{\phi}}-\frac{1-e^{4\overline{\phi}}}{K_{B}}\right )
\tilde{B}_{\varphi}-e^{4\overline{\phi}}\frac{\dot{a}}{a}\tilde{B}_{\gamma} = 0,
\end{split}
\label{persub:3}
\end{equation}
where $\theta$ denotes the fluid's velocity potential, $k = \lvert\mathbf{k}\rvert$, and $\tilde{\zeta}$ is related to $\alpha$ through Eq.
\eqref{eq:41}. To find a second equation, one may start from Eq. \eqref{eq:58}. Similar as before, I take its time derivative and use the
perturbation equations to recast the resulting expression into a more convenient form. A bit of algebra then reveals
\begin{equation}
\begin{split}
& -e^{-4\overline{\phi}}k^{2}\tilde{\Phi} + e^{-4\overline{\phi}}\left (1-\frac{\overline{\mu}}{2U}\right )k^{2}\tilde{\Psi} - e^{-4\overline{\phi}}
\frac{\overline{\mu}}{2U}\dot{\overline{\phi}}k^{2}\alpha +\left\lbrack\frac{\overline{\mu}}{U}\dot{\overline{\phi}}-3\left (\overline{\mu}\dot{\overline{\phi}} 
-\frac{\dot{b}}{b}\right )+\frac{\dot{U}}{U}\right\rbrack\left (k^{2}\tilde{\zeta} + 12\pi Ga^{2}e^{-4\overline{\phi}}\overline{\rho}\theta\right )\\
& +\left\lbrack 3\left (\frac{\dot{b}}{b}-5\dot{\overline{\phi}}\right )\frac{\dot{b}}{b} + \frac{4\pi}{U}Ga^{2}e^{-4\overline{\phi}}\overline{\rho}
-6\frac{\ddot{b}}{b}+3\left (\overline{\mu}\dot{\overline{\phi}}-\frac{\dot{b}}{b}\right )\left (\frac{\overline{\mu}}{U}\dot{\overline{\phi}}-3\left (
\overline{\mu}\dot{\overline{\phi}}-\frac{\dot{b}}{b}\right )+\frac{\dot{U}}{U}\right )\right\rbrack\tilde{\Psi}-\frac{4\pi}{U}Ga^{2}e^{-4\overline{\phi}}
\overline{\rho}\delta-3\frac{\dot{a}}{a}\frac{\dot{b}}{b}\hat{B}_{\varphi}\\
& -\frac{\dot{a}^{2}}{a^{2}}\hat{B}_{\gamma}+\left\lbrack\frac{e^{-4\overline{\phi}}}{2U}\left (\overline{\mu}k^{2}-{16\pi}Ga^{2}\overline{\rho}
\right )-3\overline{\mu}\dot{\overline{\phi}}\left (\frac{\overline{\mu}}{U}\dot{\overline{\phi}}-3\left (\overline{\mu}\dot{\overline{\phi}}-\frac{\dot{b}}{b}
\right )+\frac{\dot{U}}{U}\right )\right\rbrack\tilde{B}_{\varphi}+\left (3\overline{\mu}\dot{\overline{\phi}}-5\frac{\dot{b}}{b}-2\dot{\overline{\phi}}-
\frac{\dot{U}}{U}\right )\frac{\dot{a}}{a}\tilde{B}_{\gamma} = 0.
\end{split}
\label{persub:1}
\end{equation}
Finally, the last equation is obtained from eliminating $E$ between Eq. \eqref{persub:2} and the Hamiltonian constraint equation. Together
with the relations presented in Sec. \ref{section43}, one ends up with
\begin{equation}
\begin{split}
& -\left (2-\frac{K_{B}}{1-e^{4\overline{\phi}}}\right )k^{2}\tilde{\Phi} - K_{B}\frac{e^{4\overline{\phi}}}{1-e^{4\overline{\phi}}}k^{2}\tilde{\Psi}
+ K_{B}\left (\frac{\dot{b}}{b} - 4\frac{e^{4\overline{\phi}}}{1-e^{4\overline{\phi}}}\dot{\overline{\phi}}\right )k^{2}\alpha -2e^{4\overline{\phi}}\left (
\frac{\dot{b}}{b} - U\dot{\overline{\phi}}\right )\left (k^{2}\tilde{\zeta} + 12\pi Ga^{2}e^{-4\overline{\phi}}\overline{\rho}\theta \right ) - 8\pi Ga^{2}
\overline{\rho}\delta\\
& -2e^{4\overline{\phi}}\left\lbrack 3\left (\overline{\mu} + U\right )\frac{\dot{b}}{b}\dot{\overline{\phi}} - 3\overline{\mu}U\dot{\overline{\phi}}^{2} -
{8\pi}Ga^{2}e^{-4\overline{\phi}}\overline{\rho}\right\rbrack\tilde{\Psi}+2e^{4\overline{\phi}}\left\lbrack 3\overline{\mu}\dot{\overline{\phi}}\left (
\frac{\dot{b}}{b}-U\dot{\overline{\phi}}\right )-{8\pi}Ga^{2}e^{-4\overline{\phi}}\overline{\rho}\right\rbrack\tilde{B}_{\varphi}-2e^{4\overline{\phi}}U
\dot{\overline{\phi}}\frac{\dot{a}}{a}\tilde{B}_{\gamma}  = 0.
\end{split}
\label{persub:4}
\end{equation}
\end{widetext}

\subsection{Superhorizon limit}
\label{appendix31}
For scales well outside the horizon, the perturbation equations can be simplified by considering the limit $k\rightarrow 0$.
Taking the time derivative of Eq. \eqref{eq:58},
\begin{equation}
3\ddot{\tilde{\Phi}} - \frac{\dot{a}^{2}}{a^{2}}\hat{B}_{\gamma} - \left (\frac{\dot{a}}{a} - \frac{\dot{U}}{U} - \dot{\overline{\phi}}
\right )\left (3\dot{\tilde{\Phi}} - \frac{\dot{a}}{a}\tilde{B}_{\gamma}\right ) = \frac{ae^{-\overline{\phi}}}{2U}\dot{\gamma},
\label{persuper:1}
\end{equation}
the first step is to substitute $\dot{\gamma}$ with the help of the perturbed scalar field equations. Next, one needs to eliminate
terms involving the potential $\tilde{\Psi}$. This can be achieved by appropriately combining the time derivatives of the modified
propagation equations with the remaining perturbation equations. Finally, using Eqs. \eqref{supereq:1} and Eq. \eqref{supereq:2}
allows one to derive an equation governing the evolution of $\tilde{\Phi}$.

Defining the auxiliary quantities
\begin{equation}
\tilde{P} \equiv 3\frac{\dot{b}^{2}}{b^{2}} - 4\pi Ga^{2}e^{-4\overline{\phi}}\overline{\rho}
\label{persuper:2}
\end{equation}
and
\begin{equation}
\tilde{Q} \equiv U\left\lbrack 2\frac{\ddot{b}}{b} + \frac{\dot{b}}{b}\left (5\dot{\overline{\phi}} - \frac{\dot{b}}{b} \right )
\right\rbrack,
\label{persuper:3}
\end{equation}
the result can be expressed as
\begin{equation}
\begin{split}
&\phantom{\lbrack\times}\left (\tilde{P} - 3\tilde{Q}\right )\ddot{\tilde{\Phi}} + \tilde{P}\frac{\dot{a}}{a}\frac{\dot{b}}{b}
\hat{B}_{\varphi} + \left\lbrack\tilde{P}\left (\left (3\overline{\mu} + 2\right )\dot{\overline{\phi}} - \frac{\dot{b}}{b}
\right ) - 3\tilde{Q}\right.\\
&\left.\times\left (\left (2 + U^{-1}\right)\frac{\dot{b}}{b} + \frac{\dot{U}}{U} + \left (1 + \overline{\mu}U^{-1}\right )
\dot{\overline{\phi}}\right )\right\rbrack\dot{\tilde{\Phi}} + \tilde{Q}\frac{\dot{a}^{2}}{a^{2}}\hat{B}_{\gamma}\\
&+ \frac{\dot{a}}{a}\left\lbrack\tilde{Q}\left (2\frac{\dot{b}}{b} + \dot{\overline{\phi}} + \frac{\dot{U}}{U}\right ) +
\tilde{P}\left (\frac{\dot{b}}{b} - \overline{\mu}\dot{\overline{\phi}}\right )\right\rbrack\tilde{B}_{\gamma} = 0.
\end{split}
\label{persuper:4}
\end{equation}
Adopting the modified EdS cosmology, this eventually leads to the expression in Eq. \eqref{supereq:3}.
\bibliography{ref_short}

\end{document}